\documentclass[letter]{aa} % for the letters 
\usepackage{graphicx}
\usepackage{txfonts}
\begin{document}
   \title{Effects of dust on light-curves of $\epsilon$ Aur type stars}

%   \subtitle{I. Overviewing the $\kappa$-mechanism}

   \author{J. Budaj
%          \inst{1}
          }

   \institute{Astronomical Institute, Slovak Academy of Sciences,
              05960 Tatranska Lomnica, Slovak Republic\\
              \email{budaj@ta3.sk}
             }

   \date{Received September 15, 1996; accepted March 16, 1997}

% \abstract{}{}{}{}{} 
% 5 {} token are mandatory
 
  \abstract
   {$\epsilon$ Auriga is one of the most mysterious objects on the sky.
Prior modeling of its light-curve assumed a dark, inclined, non-transparent 
or semi-transparent, dusty disk with a central hole. The hole was necessary 
to explain the light-curve with a sharp mid-eclipse brightening.
}
   {The aim of the present paper is to study the effects of dust on 
the light-curves of eclipsing binary stars and to develop an alternative
physical model for $\epsilon$ Aur type objects which is based on 
the optical properties of dust grains.
}
   {The code Shellspec has been modified to calculate 
the light-curves and spectra of such objects. The code solves 
the radiative transfer along the line of sight in interacting binaries.  
Dust and angle dependent Mie scattering were introduced into the code for 
this purpose.
}
   {Our model of $\epsilon$ Aur consists of two geometrically 
thick flared disks: an internal optically thick disk and an external 
optically thin disk which absorbs and scatters radiation.
Disks are in the orbital plane and are almost edge-on.
We argue that there is no need for an inclined 
disk with a hole to explain the current eclipse of $\epsilon$ Aur not 
even if there is a possible shallow mid-eclipse brightening.
It was demonstrated that phase dependent light scattering and the optical 
properties of the dust can have an important effect on the light-curves 
of such stars and can even produce a mid-eclipse brightening. 
This is a natural consequence of the strong forward scattering. 
It was also demonstrated that shallow mid-eclipse brightening might 
result from eclipses by nearly edge-on flared (dusty or gaseous) disks. 
}
{}

\maketitle
%
%_______________________________________________________________

\section{Introduction}

$\epsilon$ Aur has eluded astronomers for a long time.
This star is a bright object visible by the naked eye
and known to be variable for almost two hundred years.
It is an eclipsing binary with the longest known orbital period, 27.1 yr.
A very rare eclipse which lasts for 2 years is almost over.
However, this binary star, its origin and, in particular, its secondary 
component, remain mysterious. The extremely long orbital period and 
eclipse implies that the two objects are huge.

The primary, main source of light, may be a relatively young,
massive F0Ia super-giant with the mass of about 16 $M_{\odot}$
(Stencel et al. \cite{stencel}) and radius of about 135 $R_{\odot}$
(Hoard et al. \cite{hoard}).
The secondary is unseen. Kuiper, Struve \& Str\"{o}mgren (\cite{kuiper}) 
suggested that it is a huge ($>3000R_{\odot}$) partially transparent star. 
%and most astronomy textbooks of the 20th century listed $\epsilon$ Aur as 
%the largest star in the Universe.  
Nowadays, we believe that the secondary 
is a dark and mysterious object with a radius of about 9 AU
and a mass of 13 $M_{\odot}$. 
Huang (\cite{huang}) proposed that the secondary is a dark disk seen edge on. 
Wilson (\cite{wilson}) and Carroll et al. (\cite{carroll}) argue that 
the observed sharp mid-eclipse brightening {\bf (MEB)} can only be 
explained by a tilted disk with a central opening. Ferluga (\cite{ferluga}) 
suggested that the disk is a system of rings.
\footnote{By a mid-eclipse brightening in $\epsilon$ Aur some authors
understand only a relatively sharp local maximum which appeared in 
a few eclipses near the middle of the eclipse. We propose a slightly 
more general definition of the MEB. Our MEB or Mid-Eclipse Excess 
is a convex feature near the middle of an eclipse bed. 
A common eclipse has a concave eclipse bed. 
In the pictures with reversed magnitudes it looks just the opposite.}
The main weakness of this model is that it cannot explain what 
is inside the disk and why we do not see it. Stars with enough mass are 
too bright and black holes are ruled out by the lack of X-rays.
An alternative is that the primary is an old evolved post-AGB
star with a mass of about 2.2 $M_{\odot}$ and that the secondary  
hides a main sequence B5 star with mass of about 5.9 $M_{\odot}$ in 
the center (Hoard et al. \cite{hoard}, Kloppenborg et al.
\cite{kloppenborg2010}, Takeuti \cite{takeuti} and references 
therein). 

There has been a wealth of studies during the current eclipse
at various wavelengths. Orbital solutions were recently revisited by 
Stefanik et al. (\cite{stefanik}) and Chadima et al. (\cite{chadima2010}).
Kloppenborg et al. (\cite{kloppenborg2010}, \cite{kloppenborg2011}) 
confirmed the dark disk using interferometric observations but 
they did not confirm the hole in the disk.
The spectral energy distribution was studied by Hoard et al. (\cite{hoard}).
They favor the post AGB+B5V model and concluded that the dust grains in 
the disk are unusually large (see also Lissauer et al. \cite{lissauer}).
Wolk et al. (\cite{wolk}) analyzed X-ray observations.
Recently, Chadima et al. (\cite{chadima2010}, \cite{chadima2011b}) 
questioned the presence of sharp mid-eclipse brightening and 
suggested that the photometric variability seen during eclipse is 
intrinsic to the F star.

The section below presents a motivation for this study and an outline
of important physical and geometrical effects.
These effects, dust and Mie scattering were incorporated into the code 
Shellspec. It solves radiative transfer along the line of sight in 
3D moving circumstellar environment
%Scattered light from up to two sources can be taken into account
%assuming optically thin environment
(Budaj \& Richards \cite{budaj2004}, see also Budaj \cite{budaj2011},
Miller et al. \cite{miller}, Tkachenko et al. \cite{tkachenko}, 
Chadima et al. \cite{chadima2011a} for some modifications and applications).
In the next sections we apply this code to $\epsilon$ Aur.

\section{Motivation}
\label{mot}

The standard explanation of the MEB relies on the presence of a disk
and pure geometry. It requires a chain of strict assumptions.
The disk is optically thick but geometrically thin. 
It has to be inclined with respect to the orbital plane, and must 
contain a central hole. The hole may not be present all the time 
which requires a lot of energy to fill it in or to open it again 
since it resides in a region of strong gravitational field. 
Moreover, this does not explain everything and raises the question 
why we do not see the object in the center.

\begin{figure}
\centering
\includegraphics[angle=-90,width=8.3cm]{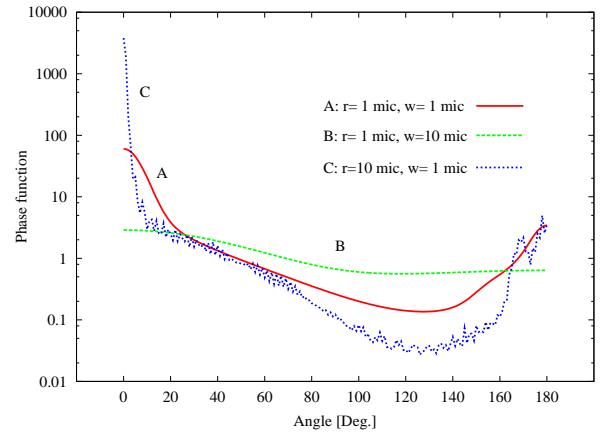}
\caption{
Dust phase function as a function of angle for forsterite. 
Calculated for spherical particles with radii, $r$, equal to 1 and 
10 micron for wavelengths, $w$, equal to 1 and 10 micron. 
Note the strong forward scattering peak near 0 degrees and less 
pronounced backward scattering peak near 180 degrees for larger grains
at shorter wavelengths.
}
\label{f1}
\end{figure}

There might be an alternative physical explanation of the MEB.
It is natural that there is a smooth transition from 
the optically thick to optically thin disk until it eventually vanishes.
The optical behavior of dusty medium can be described by the Mie theory
and characterized by the microphysical properties of dust gains. 
These are the grain size or size distribution, complex refractive index
or chemical composition, shape of grains, etc..
The striking property of dust grains is that they do not scatter 
light isotropically but exhibit strong forward scattering, 
less pronounced backward scattering and weak side scatter.
This dependence of scattered light on the angle between
the dust grain, source of light, and the observer is referred to as
a phase function.
Fig. \ref{f1} illustrates the behavior of the dust phase function
of forsterite which constitutes one of the most refractory, abundant,
and opaque dust grains.
Notice the strong forward scattering for angles close to zero.
It is most pronounced for larger grains and at shorter wavelengths.
Backward scattering for angles close to 180 degrees is not that pronounced
but has a similar behavior. 
These phase functions, scattering and absorption opacities were calculated 
using a Mie scattering code by Kocifaj (\cite{kocifaj04}) and 
Kocifaj et al. (\cite{kocifaj08}). The complex index of refraction of 
forsterite was taken from J\"ager et al.(\cite{jager}).
To suppress a ripple structure which would appear in 
the phase function of spherical mono-disperse particles, 
the poly-disperse Deirmendjian (\cite{deirmendjian}) distribution of 
particle sizes was assumed.

Let us assume a hypothetical 'eclipsing binary star' which consists
of a star (source of light ) and a distant ball of optically thin dust.
Both objects revolve around their center of mass.
The dusty ball will scatter the light in the forward direction creating  
a beam of light analogous to the light-house effect.
Whenever the beam hits the observer, which happens mainly during 
the eclipse, a pulse will be detected.
The dusty ball may also cause attenuation of the light from the source
during the eclipse. These two effects will compete. 
If the main source of light disappears, e.g. during the total eclipse 
by some optically thick object, then scattering by the optically 
thin dust may completely dominate the observed radiation.
Consequently, such a 'naked' light-house effect might be a natural 
explanation of the MEB observed in some eclipsing binary stars.

Apart from the above mentioned effect we propose that the shallow
MEB may also result from edge-on flared disks.
Suppose that the disk is flared, homogeneous and optically 
thin (or at least partially transparent). 
Then the edges of the disk may have a larger effective 
cross-section and optical depth along the line of sight than 
the central part of the disk and consequently might attenuate 
the stellar light more effectively before and after the mid-eclipse.
This effect relies mainly on the shape of the disk and opacity and 
is independent on the source of opacity which is why it could work for 
both dusty and gaseous disks.

\section{Observations}
\label{obs}
Observations which were used for comparison with our calculations
were taken from the AAVSO database (Henden \cite{henden}). They were obtained 
by many observers who contributed to the database during the current eclipse 
of  $\epsilon$ Aur. Only the observations in the V filter were considered 
here. These observations indicate the presence of a shallow mid-eclipse
brightening but it is not as sharp and pronounced as some might have 
anticipated.  We fitted an eclipse bed (phases $<-0.19,0.19>$) with 
a quadratic function $y=a+bx+cx^{2}$ (see Fig.\ref{f3}) and obtained 
the following coefficients:
$a=3.690\pm0.004, b=-1.1\pm0.2, c=140\pm20$. The $c$ coefficient is
positive which indicates that the eclipse bed has a convex shape and 
thus represents a mid-eclipse brightening. Its error indicates that 
its significance is beyond 5 $\sigma$. Notice that if $b\neq0$ then maximum 
of the function is not at the middle of the eclipse. However, the quadratic 
term, which constitutes the MEB, still has maximum at the center of 
the eclipse. This justifies a need for a more general definition
of the MEB or a Mid-Eclipse Excess. Presence of the linear term, longer 
ingress and steeper egress indicate that disk is not perfectly symmetric 
but suffers from some disturbance. Its leading part might be more 
extended or disk slightly inclined out of the orbital plane (warped?).

\section{Light-curve of $\epsilon$ Aur}

In this section we will carry out calculations of the light-curves
and study effects outlined in Sect.\ref{mot}.
If not mentioned otherwise we will follow the alternative model
of $\epsilon$ Aur (Hoard et al. \cite{hoard}) and assume that the main 
source of light is a star with radius of 135 $R_{\odot}$, 
$T_{\rm eff}=7750$K, and mass of 2.2 $M_{\odot}$. 
A quadratic limb darkening for filter V was 
applied to it with coefficients 0.38, 0.28 (Claret \cite{claret}).
Its spectrum was approximated by a black-body.
The second object is separated by 18.1 AU, has a mass of 
5.9 $M_{\odot}$, and is enshrouded in a disk.
Light-curves were computed for 550 nm and a polydisperse 
distribution of particle sizes with a typical radius of 4 $\mu$m.

\subsection{Effect of forward scattering on the eclipse}

\begin{figure}
\centering
\includegraphics[angle=-90,width=8.3cm]{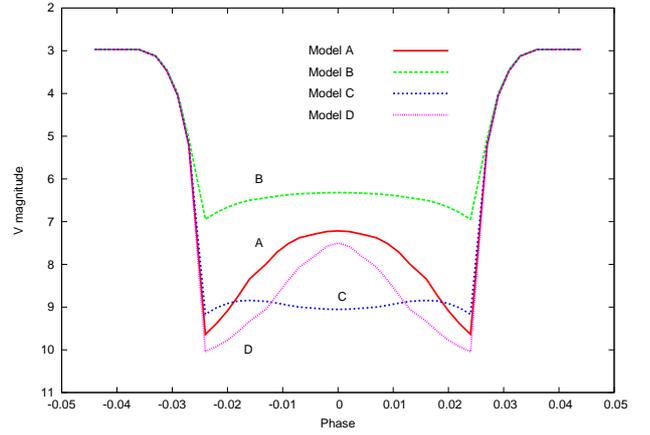}     
\caption{
Hypothetical total eclipse of $\epsilon$ Aur by a dark, geometrically and
optically thick, edge-on disk of dust. This disk causes the total eclipse 
and does not transmit any light during the eclipse. 
Apart from this non-transparent disk our model includes optically 
thin dust regions of different shapes:
Model (A) has a spherical shell of dust,
Model (B) has a flat disk region, 
Model (C) has a flared disk, and 
Model (D) has a sort of jet-like region. 
These regions scatter the light toward the observer which is 
seen during the total eclipse. 
Because of the strong forward scattering this scattered light may be
most intense in the middle of the eclipse and might give rise to 
a mid-eclipse brightening. See the text for more details.
}                                              
\label{f2}               
\end{figure}

This section illustrates the pure effect of forward scattering
on the light-curve of $\epsilon$ Aur during the 
{\it hypothetical total eclipse}.
We assume that the main source of light is eclipsed by an opaque dark 
and geometrically thick disk seen edge-on. To observe a 'pure 'forward 
scattering effect we assume that the eclipse is total i.e. the star is 
completely hidden behind the optically and geometrically thick disk. 
Calculations are presented in Fig. \ref{f2}.
The opaque disk has the shape of a slab with radius of 730 $R_{\odot}$ and 
thickness of 272 $R_{\odot}$. Apart from the opaque disk we included 
a few different optically thin dust components which scatter the light
toward the observer: 
Model (A) has a sherical shell of dust with a radius of 
400 $R_{\odot}$ and density of $5\,10^{-18} \rm g\,cm^{-3}$;
Model (B) has a slab of dust above and below the opaque disk 
(sort of atmosphere) which is 250 $R_{\odot}$ thick with density of 
$5\,10^{-18} \rm g\,cm^{-3}$;
Model (C) is a flared disk. The flared disk in our model has the shape 
of a rotating wedge with an opening angle. In this case we assumed 
an opening angle of 50 degrees and density of $0.8\,10^{-18} \rm g\,cm^{-3}$;
Model (D) has the shape of a bipolar outflow (or a jet perpendicular to 
the opaque disk) with half-opening angle of 30 degrees, extending from 
the center up to 400 $R_{\odot}$, and dust density of 
$2\,10^{-17} \rm g\,cm^{-3}$.
One can indeed observe the above mentioned MEB in most cases. 
Interestingly, a flared disk may scatter more light before and 
after the mid-eclipse. 
Jets or structures with central dust concentrations may provide 
quite sharp MEB similar to that claimed in a few past eclipses of 
$\epsilon$ Aur.
In general, the shape of the light-curve or spectral lines
would be given by the shapes of the opaque and non-opaque objects, 
geometry of the eclipse, wavelength, properties of the dust and gas such 
as its density, temperature, chemical composition, velocity field, and 
grain size. In this configuration the amount of scattered light can reach 
a few percent but it is heavily dependent upon the parameters mentioned 
above, particularly on those which determine the width of the forward 
scattering peak and angle between the scattering object, source of light, 
and observer.

\subsection{Effect of flared edge-on disk on the eclipse}

\begin{figure}
\centering
\includegraphics[angle=-90,width=8.8cm]{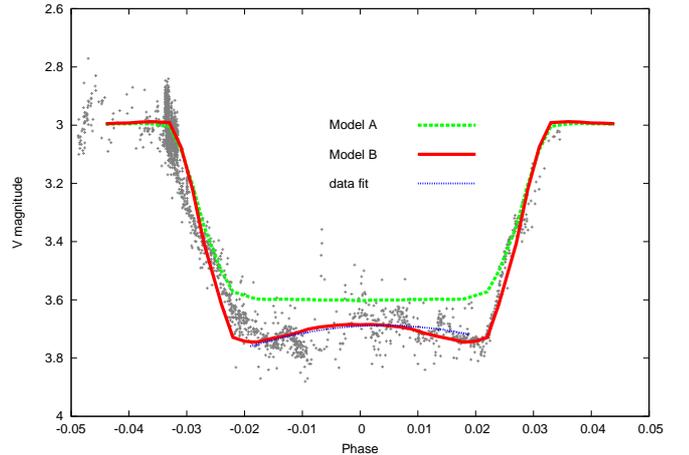}     
\caption{
Eclipse of $\epsilon$ Aur by a dark, geometrically thick, flared
disk of dust. The disk consists of two parts:
(1) The flared optically thick part which causes most of the eclipse, and 
(2) a flared optically thin part which causes additional absorption,
scattering and mid-eclipse brightening. 
Model (A): Disk has only part (1). 
Model (B): disk has both parts part (1) and part (2).
Mid-eclipse brightening arises mainly because the edges of the flared disk 
are more effective in the attenuation of the stellar light than the central 
parts of the disk. Dotted line is a best fit quadratic function to the
eclipse bed.
Crosses - observations from AAVSO (Henden \cite{henden}). 
}                                              
\label{f3}               
\end{figure}

The previous section demonstrated the effect of a pure forward scattering
during the {\it hypothetical total eclipse of $\epsilon$ Aur}.
However, in reality, the disk may not be thick enough to cause the total 
eclipse and the eclipse will be only partial. Our first experiments with 
disks of various shapes revealed that if the eclipse is partial and the disk 
has the shape of a slab then it is very difficult to reproduce the observed 
shallow MEB.
In Sect.\ref{mot} we suggested another alternative that optically thin 
but geometrically thick flared disks might also produce a shallow MEB. 
That is why in this section we study partial eclipses by flared disks. 
The best fit to the observations (not necessarily the only possible solution)
had the following geometry and components:
(1) an optically and geometrically thick flared disk with an opening angle 
of about 7.5 degrees and radius of 690 $R_{\odot}$ (3.2 AU), immersed in 
(2) an optically thin and geometrically thick flared disk with an angle 
of 28 degrees and radius of 690 $R_{\odot}$ and density of 
$6\,10^{-18} \rm g\,cm^{-3}$.
Both disks were located in the orbital plane and were almost edge-on
with the inclination of the orbital plane of 89.1 degrees.
Synthetic light-curves are compared to the observations (V band) in 
Fig.\ref{f3}.
The dust phase function was convolved with the normalized limb 
darkened profile with the full width corresponding to the angular 
diameter of the F star as seen from the centre of the disk.
In this way we took into account the finite dimension of the source 
of light also in the calculations of the scattering.

%zvaz ci toto tam davat??
We have demonstrated the effect of forward scattering and a flared disk
on the eclipse of $\epsilon$ Aur. However,
there are other objects in which this mechanism may be important.
Pre-main sequence stars often contain dusty accretion disks. 
%KH 15D is a recent example of such a star with a mid-eclipse brightening 
%(Kearns \& Herbst \cite{kearns}).
%Agol et al. (\cite{agol}) suggested that a substantial fraction of 
%the light during eclipse is scattered off dust grains.
%Chiang \& Murray-Clay (\cite{chiang2004}) proposed a model with 
%a circumbinary ring to explain the complex variability of the star.
%T Tau and protostellar envelopes were modeled using 3D and 2D radiative 
%transfer and Henyey-Greenstein phase function by Wood et al. (\cite{wood})
%and Whitney et al. (\cite{whitney}). 
%Dust scattering in other interesting
%T Tau stars was studied by Close et al. (\cite{close}) and 
%Voshchinnikov \& Karjukin (\cite{voshchinnikov}).
Evolved objects, symbiotic Miras and/or AGB stars may also contain 
dusty regions of various shapes where phase dependent Mie scattering 
might play an important role.
Vinkovi\'{c} et al. (\cite{vinkovic2004}) studied bipolar outflow structures 
in the dusty winds of AGB stars by 2D radiative transfer with isotropic 
scattering. Kotnik-Karuza et al. (\cite{kotnik}) studied the properties of 
the circumstellar dust during obscuration events in the symbiotic Miras.
Kudzej (\cite{kudzej}) mentioned seventeen other eclipsing binary systems
with mid-eclipse brightening and suggested that it is due to the refraction 
in the atmospheres of stars.
%de Kok et al. (\cite{kok}) studied the influence of non-isotropic scattering 
%of thermal radiation on spectra of brown dwarfs and hot exoplanets.

\section{Conclusions}

A new alternative model of the $\epsilon$ Aur type stars was
suggested.
This model is based on the optical properties of dust grains.
It assumes that dust concentrates in two regions: an optically thick
disk and optically thin regions of different shapes.
It was demonstrated that an optically thick flared disk coated with
an optically thin layer can reproduce the current eclipse observations of 
$\epsilon$ Aur very well. 
The model takes into account Mie scattering on dust, i.e.
extinction due to the absorption and scattering, as well as
thermal and scattering emission.

It was demonstrated that angle dependent Mie scattering is important
in such systems. Under certain circumstances, mainly if the dark dust
is eclipsing a strong source of light, the forward scattering on dust
can be crucial and can even produce a mid-eclipse brightening.
It was also demonstrated that near edge-on flared disks can produce
eclipses with a shallow mid-eclipse brightening and that this might 
be an explanation of the shallow mid-eclipse brightening feature of 
$\epsilon$ Aur.
 
This model provides an alternative to an inclined disk with a central 
hole invoked to explain systems with mid-eclipse brightening.
It does not raise the question why we do not see the object in 
the center of the disk. We may not see it because in our model 
it may be obscured by the disk.
Potential variability in the mid-eclipse brightening might be caused
by subtle changes in the remote, low density, optically thin regions of dust
located at small gravitational potential, or by small changes in 
the inclination (or warping) of the disk and does not have to involve dense
matter at high gravitational potential in the central hole of the disk.

\begin{acknowledgements}
We acknowledge with thanks the variable star observations from the AAVSO
International Database contributed by observers worldwide and used in this
research. We would like to thank Dr. Kocifaj for his help with his code
and comments on the manuscript and an anonymous referee, 
Profs. P. Harmanec, M. Richards, S. Rucinski, and B. Kloppenborg for 
their comments on the manuscript.
This work has been supported by the VEGA grants of the Slovak Academy of
sciences Nos. 2/0074/09, 2/0078/10, 2/0094/11.

\end{acknowledgements}

\end{document}